**Key Points:**

- A comprehensive study of the plasma density in the magnetotail lobes is presented
- The lobe density is asymmetric; the density is higher in the northern lobe
- North-south differences in ion outflow are the most probable explanation for the asymmetry

**Correspondence to:**
S. Haaland,
stein.haaland@uib.no

# North-south asymmetries in cold plasma density in the magnetotail lobes: Cluster observations

**S. Haaland[1,2], B. Lybekk[3], L. Maes[4], K. Laundal[1], A. Pedersen[3], P. Tenfjord[1], A. Ohma[1], N. Østgaard[1], J. Reistad[1], and K. Snekvik[1]**

[1]Birkeland Centre for Space Science, University of Bergen, Bergen, Norway, [2]Max-Planck Institute for Solar Systems Research, Göttingen, Germany, [3]Department of Physics, University of Oslo, Oslo, Norway, [4]Royal Belgian Institute for Space Aeronomy, Brussels, Belgium

**Abstract** In this paper, we present observations of cold (0–70 eV) plasma density in the magnetotail lobes. The observations and results are based on 16 years of Cluster observation of spacecraft potential measurements converted into local plasma densities. Measurements from all four Cluster spacecraft have been used, and the survey indicates a persistent asymmetry in lobe density, with consistently higher cold plasma densities in the northern lobe. External influences, such as daily and seasonal variations in the Earth's tilt angle, can introduce temporary north-south asymmetries through asymmetric ionization of the two hemispheres. Likewise, external drivers, such as the orientation of the interplanetary magnetic field can set up additional spatial asymmetries in outflow and lobe filling. The persistent asymmetry reported in this paper is also influenced by these external factors but is mainly caused by differences in magnetic field configuration in the Northern and Southern Hemisphere ionospheres.

## 1. Introduction

In terms of volume, the magnetotail lobes bound by the plasma sheet on one side and the magnetopause on the other side comprise a major part of the magnetotail. The magnetic field is primarily directed sunward (northern lobe) or antisunward (southern lobe) with only small northward and dawn-dusk components. One end of the magnetic field lines of the lobes connects to the solar wind downtail, and the other end connects to the polar cap of the ionosphere. The plasma density is very low, and the plasma is predominantly of ionospheric origin [e.g., *Chappell et al.*, 1987, 2000], though there is also some transfer of plasma from the solar wind across the magnetopause [e.g., *Hultqvist et al.*, 1999; *Shi et al.*, 2013, and references therein]. Filling of lobe flux tubes starts with ionization of atoms and gas molecules in the thermosphere, accompanied by upflow due to thermal and electromagnetic forces. Some of this material can reach escape velocities and be further accelerated and eventually evacuated into space. Transport of the outflowing ionized material is largely governed by electromagnetic forces and the transport path from the ionosphere to the nightside magnetosphere go through the magnetotail lobes. Much of the variability in outflow is therefore reflected as variations in plasma density in the magnetotail lobes. The magnetotail lobe regions thus constitute important interfaces between plasma sources in the solar wind and ionosphere on one side and the nightside magnetosphere and tail plasma sheet on the other side.

Measuring the plasma population in the polar cap and lobe regions using traditional particle instruments onboard satellites is challenging. Spacecraft charging prevents ions with energies below the electric potential of the spacecraft from reaching the detectors. Electron detectors, on the other hand, are often significantly contaminated by photoelectrons. In addition, the low densities encountered in this region usually require long integration times and/or large geometric factors of the sensor to obtain sufficient count rates to derive reliable plasma moments. Measurements based on active and passive sounders, such as the Wideband Plasma Wave Investigation [see *Gurnett et al.*, 1997] and WHISPER [see *Trotignon et al.*, 2001] (Waves of HIgh frequency and Sounder for Probing of Electron density by Relaxation) flown on Cluster, in which plasma densities are estimated from plasma resonance lines or wave power cutoff frequencies, can provide very accurate local plasma density measurements. However, these instruments and methods also have limitations. It is not always possible to identify discrete, unambiguous resonance lines in the wave spectra using automated procedures. Manual inspection is time consuming and often not practicable for large data sets. Limitations in instrument frequency range also constrain the plasma density range possible to detect. Even with reliable detection of resonance lines, these measurements can be skewed by photoelectron emissions from the spacecraft surface [e.g., *Szita et al.*, 2001].

An alternative approach to measure cold plasma density has therefore been to derive densities based on the spacecraft potential. The basic technique and initial calibration effort goes back to *Knott et al.* [1984] and was later refined by *Schmidt and Pedersen* [1987]. Both of these studies used data from geostationary orbit, where spacecraft potentials were up to a few tens of volts. Later, the methodology was also applied in other regions, such as the lobe and polar regions [*Pedersen*, 1995; *Escoubet et al.*, 1997a; *Ishisaka et al.*, 2001; *Laakso*, 2002; *Pedersen et al.*, 2008; *Lybekk et al.*, 2012] where even higher spacecraft potentials are frequently encountered. This technique is robust, easy to implement, and useful also for large data sets but requires careful calibrations.

A comprehensive study of the lobe density using Cluster data from the years 2001–2007 was conducted by *Svenes et al.* [2008] using improved calibration based on efforts by *Pedersen et al.* [2001]. They reported density values ranging from 0.007 to 0.5 $cm^{-3}$ (where 0.5 was the upper limit of their calibrations), with a typical density value around 0.05 $cm^{-3}$. *Haaland et al.* [2012] performed a similar study focusing on polar cap densities, with particular focus on variations due to solar illumination and variations due to solar wind-magnetosphere interactions and geomagnetic activity.

In this paper, which is partly a follow up on the *Svenes et al.* [2008] and *Haaland et al.* [2012] studies, we utilize a larger data set from Cluster, spanning from 2001 until 2016 and with improved calibrations [*Pedersen et al.*, 2008; *Lybekk et al.*, 2012] and a larger density range. The much larger data set allows us to focus on two new science issues: spatial asymmetries in the cold plasma distribution, in particular interhemispheric differences in cold plasma density in the lobe regions. This paper is organized as follows. In section 2 we provide a brief overview of the data and the methodology used to infer density from spacecraft potential. In section 3, we describe the analysis procedure in some detail. Results are presented in section 4. Finally section 5 provides a summary and discussion of the results.

## 2. Methodology: Deriving Plasma Density From Spacecraft Potential

When a spacecraft is exposed to solar illumination, photoelectrons are emitted from the spacecraft surface material [*Feuerbacher and Fitton*, 1972]. This loss of electrons can be regarded as a current from the ambient plasma *into* the spacecraft. As a consequence, the spacecraft potential will increase until this current is balanced by a similar current *out* of the spacecraft. An equilibrium condition with balanced currents will arise almost immediately, but the spacecraft potential relative to the ambient plasma will vary as the spacecraft traverses different plasma regimes. This property can be utilized to derive local plasma densities through a functional relation between spacecraft potential, $V_{SC}$ and the ambient electron density, $N_e$:

$$N_e = Ae^{-V_{SC}/B} + Ce^{-V_{SC}/D} \tag{1}$$

This relation is based on models of photoemissions and a number of measurements from Cluster. We refer to *Pedersen et al.* [2008] for details. $A$, $B$, $C$, and $D$ are calibration factors discussed below. Assuming quasi-neutrality, the derived electron density is identical to the ion density, and we will simply use the term "plasma density" hereafter.

An electric field experiment on a spinning satellite like Cluster, using spherical double probes on long booms, can measure electric fields and, as a byproduct, provide information about spacecraft potential. This is possible because the probes are operated with bias currents, bringing them close to the local (ambient) plasma potentials near the probes, resulting in better probe-plasma impedance and providing a good reference for the spacecraft potential.

In reality, there are also some photoemissions from spacecraft peripheries such as booms and antennas. On Cluster, the probes of the Electric Field and Wave experiment [see *Gustafsson et al.*, 2001] are mounted on approximately 40 m long wire booms. The experiment has been designed to reduce the influence of photoelectrons from the boom tips by extending the probes from the boom tip and the preamplifier on 1.5 m long, very thin wires. The number of photoelectrons from this thin wire is very small compared to photoelectrons from a spherical probe, and more accurate measurements can thus be achieved. Cluster electric field data, obtained over many years, have demonstrated an improved data quality and have made it possible to get more reliable spacecraft potential measurements than earlier missions.

The calibration coefficients *A*, *B*, *C*, and *D* in equation (1) are determined from in-flight calibrations and implicitly contain information about spacecraft surface properties, plasma temperature, illuminated surface area, and any changes in solar illumination. These parameters, in particular solar irradiance, and consequently the relation between potential and density change with time. Also, surface properties can be different for different spacecraft, even between the four Cluster spacecraft. A few years ago, *Lybekk et al.* [2012] performed a careful cross calibration between the above method and the WHISPER and the Plasma Electron and Current Experiment [see *Johnstone et al.*, 1997] on board the Cluster satellites and were able to determine the calibration factors for all spacecraft valid over a wide range of solar activity. In this paper, we utilize a large data set of density measurements based on the above method. Details about the calibration coefficients and their change over time are given in *Lybekk et al.* [2012, Appendix A1].

### 2.1. Error Estimates, Data Processing, and Caveats

A discussion of measurements uncertainty was already given by *Haaland et al.* [2012] with a more comprehensive discussion in *Lybekk et al.* [2012]. The measurement uncertainty for a single *Ne* measurement depends on several factors. Equation (1) assumes that the measured $V_{SC}$, which is the measured difference in voltage between the spacecraft body and the boom mounted probes, is the spacecraft voltage relative to the ambient plasma. In reality, the probes themselves have a small potential, about $2 \pm 0.5$ V [*Pedersen et al.*, 2008]. Since the measurements reported in this paper are taken in the polar cap and lobe regions, the spacecraft potential is typically several tens of volts. The uncertainty in probe potential will therefore typically contribute an uncertainty of less than 5%. Additional uncertainties arise form the determination of the calibration coefficients in equation (1). Overall, *Lybekk et al.* [2012] estimated that the uncertainty of an individual density measurement is of the order of ± 20%.

All Cluster satellites have an active spacecraft control (ASPOC) [see, *Torkar et al.*, 2001] which emits a beam of indium ions having the effect of reducing the positive spacecraft potential to a few volts. ASPOC was (nominal operation ended in 2003) operated during short planned periods in order to obtain information about low-energy electron and ion spectra. Measuring electron density based on spacecraft potential measurements cannot be used during these periods. Cluster also has an active electric field drift experiment (EDI) [see *Paschmann et al.*, 2001; *Quinn et al.*, 2001] which emits currents in the form of an electron beam. Current imbalances caused by EDI can be compensated by proper calibration [see *Lybekk et al.*, 2012] if the emitted currents are small, but for higher EDI currents, equation (1) does not hold, and no reliable density measurements can be made. Periods of ASPOC operation and high EDI currents are flagged in the spacecraft housekeeping data and have been removed from our data set.

Cluster data used in this study are 1 min averages, where each 1 min average is calculated as the median value of 15–16 spins (the Cluster spin period is approximately 4 s). Further details about the processing can be found in *Lybekk et al.* [2012]. All measurements are taken from the Cluster Science Archive [*Laakso et al.*, 2010]. Auxiliary data such as solar wind measurements and geomagnetic activity indices are also 1 min averages from CDAWeb. The data are stored as tables in a database, and scripts with structured queries are used to extract subsets of the data and for filtering.

### 2.2. Defining the Lobe

Conceptually, the lobes are defined as the region of the magnetotail where magnetic field lines are open to the solar wind at one end. In practice, we have to identify this region either from characteristic plasma properties [e.g., *Vaisberg et al.*, 1996; *Koleva and Sauvaud*, 2008] or magnetic field properties alone [e.g., *Coxon et al.*, 2016]. In this study, we identify the lobes using a combination of spacecraft potential, plasma beta, and magnetic field orientation. The definition largely follows the same classification as *Svenes et al.* [2008] but using magnetic field strength and direction rather than fixed positions in space to distinguish between northern and southern lobes.

To investigate north-south asymmetries in the density, we first need to define “northern” and “southern” lobes. Noting that the magnetotail configuration changes significantly both on a diurnal and seasonal timescale, and also moves significantly (tail flapping and flaring) in response to changes in the solar wind, a simple classification according to location only would be misleading. We therefore use the magnetic field direction to determine whether we are in the northern or southern lobe; $B_X$ should always points sunward in the northern lobe and tailward in the southern lobe. Also, the $B_X$ component of the magnetic field should be dominant, and the magnetic pressure should be much higher than the thermal pressure (low plasma beta). As we will see below, the main results of this paper are not very sensitive to changes in these parameters. This is reassuring, in particular since the plasma beta can usually not be accurately determined in the tenuous plasma of the lobes. Table 1 shows our definition of the lobes.

**Table 1.** Filter Criteria Used to Define the Lobes and Amount of Observations Satisfying These Conditions[a]

| A | B<br>$B_X$ (nT) | C<br>$V_{SC}$ (V) | D<br>Ne ($cm^{-3}$) | E<br>$\beta$ | F<br>Coverage (h) |
|---|---|---|---|---|---|
| Northern lobe | $\geq 5$ | $\geq 25$ | $\leq 1.0$ | $\leq 0.05$ | 8499 |
| Southern lobe | $\leq -5$ | $\geq 25$ | $\leq 1.0$ | $\leq 0.05$ | 59086 |

[a] $B$ and $B_X$ are the magnetic field magnitude and the magnetic field along $X_{GSE}$, respectively; $V_{SC}$ is the spacecraft potential, and $\beta$ is the ratio between thermal particle pressure and magnetic pressure. Only nightside observations ($X_{GSE} \leq 0$) are included. Due to the evolution of the Cluster orbit, with apogee gradually shifting southward, we have more lobe measurements from the southern lobe.

To obtain unbiased, quantitative estimates of the lobe density and thus be able to assess whether there are any spatial asymmetries, we will further constrain the above selection criteria to only include data from certain regions (either in terms of location or in terms of magnetic flux tube volume).

### 2.3. Data Set Characteristics

Cluster's prograde 57 h polar orbit means that the satellites spend significant time in the low density, polar cap, and lobe regions of the magnetosphere where the technique in the above section can be utilized. In *Svenes et al.* [2008], only data from Cluster SC4, where the EDI instrument was not operating, could be used. The new calibration by *Lybekk et al.* [2012] takes currents from the EDI instrument (at least up to a certain beam current) into consideration, so data from all four Cluster satellites can be used. The large amount of data collected by the Cluster spacecraft over 16 years allows us to address the question of spatial asymmetries in cold plasma density. Using our definitions of the lobe, i.e., nightside locations with a $B_X$ dominated magnetic field and low plasma beta — see details in Table 1 — means that approximately 67'000 h of data from the tail lobes spanning more than a full solar cycle are available. Due to the Cluster orbit, with apogee gradually moving southward, coverage is significantly better in the Southern Hemisphere, in particular during later years.

Cluster's apogee is in the tail from late July to around October and traverses the midtail (noon-midnight meridian) around equinox. This means that the bias due to different solar illumination of the northern and southern ionosphere is small near the noon-midnight plane. On the other hand, this orbit also means that the Cluster data set is not well suitable to assess dawn-dusk asymmetries in lobe density, as these will largely be results of seasonal variations in solar illumination.

Figure 1 shows an overview of the full data set — more than 14 million 1 min records — projected onto the *XZ*, *XY*, and *YZ* GSE planes, respectively. The evolution of the Cluster orbit is indicated in Figure 1a with sketches of typical orbits in the early (2001) and late (2016) phase of the Cluster mission. The corresponding seasonal evolution is indicated in Figure 1b where orbits for late July, equinox, and early November are indicated. This plot gives an overview of the data coverage, but note that measured density values and data coverage are strongly influenced by the orbit of Cluster. At the time of writing (mid-2016) not all auxiliary data (i.e., plasma moments needed to determine, e.g., plasma beta) had been ingested into the Cluster Science Archive. There are also data gaps due to spacecraft maneuvers, eclipses, and temporary shutdowns of instruments. The measurements are therefore neither complete nor continuous in time. In the subsequent sections, we will use subsets of this full data set to address specific science questions.

Distributions, shown in Figure 2, provide further insight into the data, now only containing measurements from the magnetotail lobes as defined in Table 1. Figures 2a to 2c show density distributions, radial distributions, and density versus radial distance. Blue color indicate northern lobe values; red color show southern lobe values. Dashed lines in Figures 2a and 2c show the corresponding densities but based on observations taken within 15 days around equinox only. Seasonal difference due to asymmetries in illumination and thus asymmetries in ionospheric outflow from the two hemispheres should then be largely eliminated. There is no tilt bias in our data set, and daily variations in lobe density due to dipole tilt are negligible.

From Figure 2a, one notes non-Gaussian distributions with significant spread for both hemispheres. This applies both to the full data set and for the equinox data set (this is also typically the case for any further subsetting discussed below). Statistical moments like mean, median, and mode will be different, and they are often comparable to the standard deviation of the distributions. Unless otherwise stated, we will give median values as a measure of averages to avoid strong influences from single events and outliers (i.e., tail of distribution). We also calculated standard error, $S = \sigma/\sqrt{N}$, where $N$ is the number of samples in a distribution, and $\sigma$ is the standard deviation of the distribution. Due to the large number of samples, the standard error is typically very small and therefore not shown in the plots.

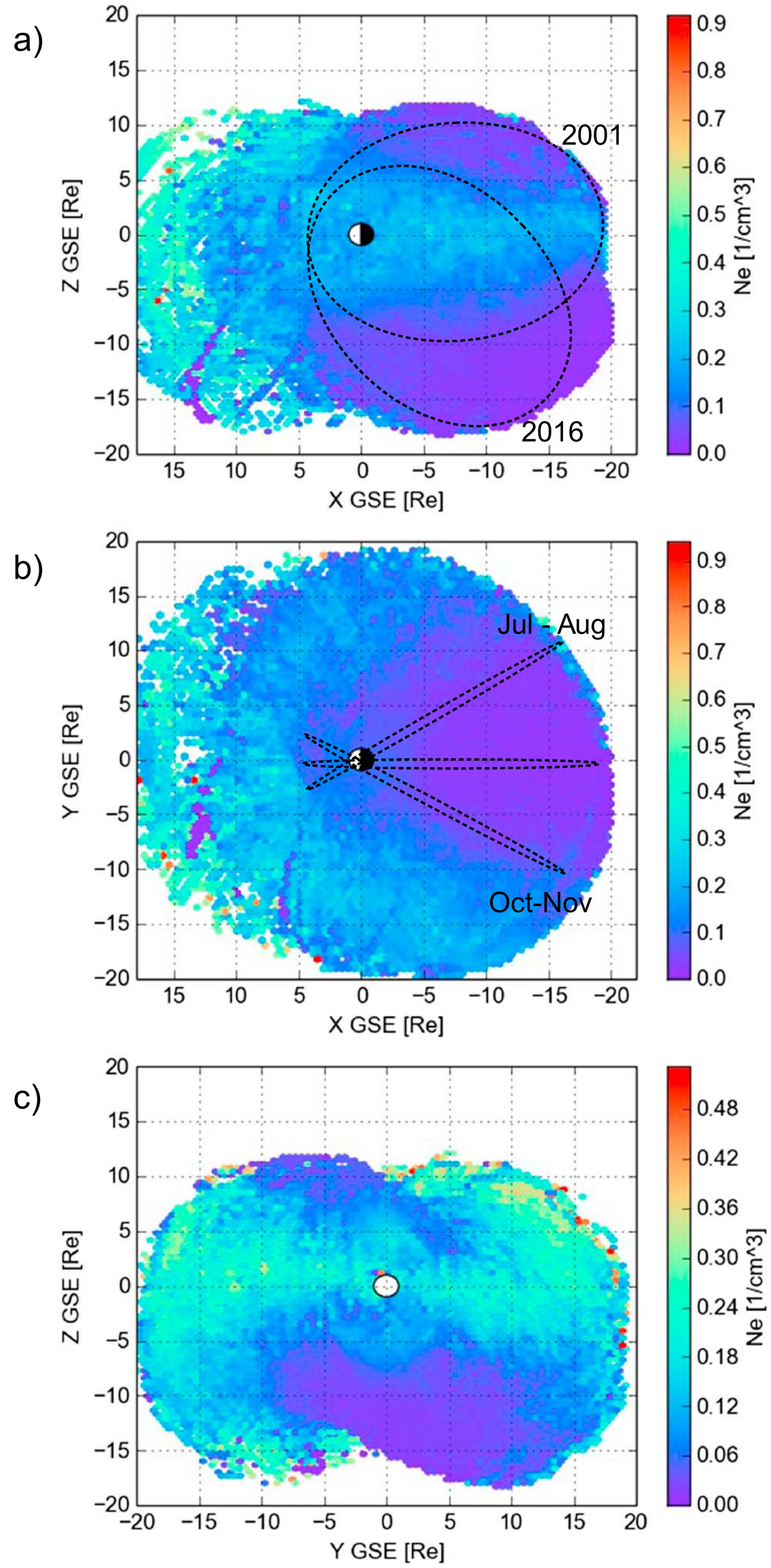

**Figure 1.** Overview of *all* Cluster density measurements based on spacecraft potential measurements for the years 2001–2016. Colors indicate absolute density, averaged into approximately $0.2 \times 0.2\ R_E$ bins. (a) $XZ_{GSE}$ projection seen from dusk toward dawn. Typical orbits for the early years (2001) and recent years are indicated. (b) $XY_{GSE}$ projection as seen from north down to the ecliptic plane. The seasonal evolution of the orbit is indicated. Cluster has apogee in the midtail around equinox. (c) $YZ_{GSE}$ projection as seen from the Sun.

In addition to the peaks around 0.07 cm$^{-3}$ for northern lobe and 0.05 cm$^{-3}$ for Southern Hemisphere, we also note a second set of plateaus or maxima above 0.1 cm$^{-3}$ for both hemispheres—also with a north-south asymmetry. This is possibly indicating a second plasma population originating from a different source ([see e.g., *Hirahara et al.*, 1996] cusp/cleft rather than polar cap), but a detailed investigation is beyond the scope of this paper. Alternatively, there is still some influence from cold plasma near the plasma sheet and boundary layers [e.g., *Seki et al.*, 2003]. When we do subsetting of the full data set, these second maxima become less pronounced or disappear, though.

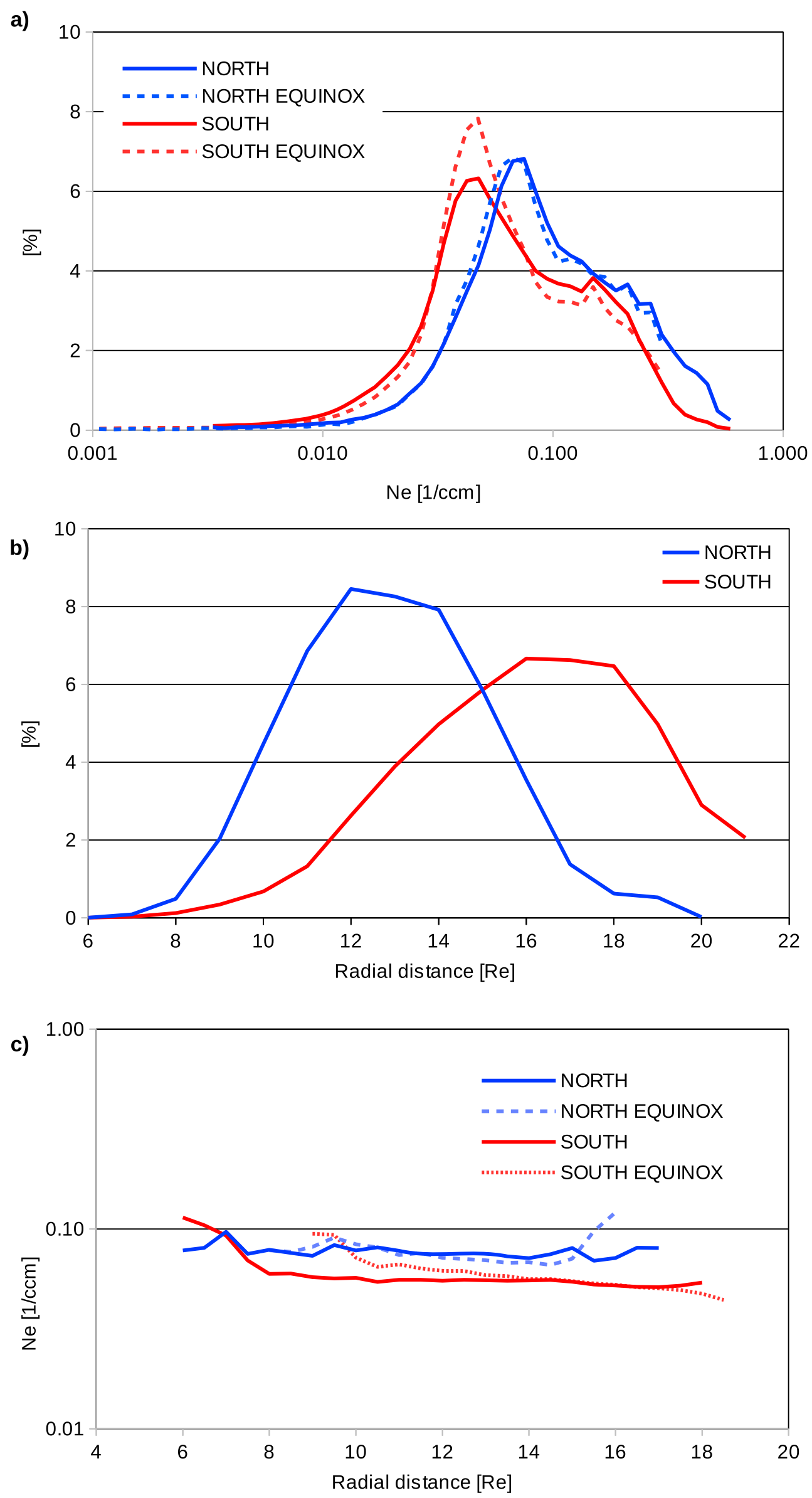

**Figure 2.** Characteristic distributions of the data set. (a) Density distribution. (b) Radial distribution of the measurements. (c) Density versus radial distance. Blue color indicates Northern Hemisphere values; red colors indicate Southern Hemisphere. Dashed lines indicate values taken within 15 days of equinox. Only lobe values, i.e., values satisfying the criteria in Table 1, are included in these distributions.

Figure 2b shows the radial distribution of measurement locations for the two lobes. One here notes a bias due to the evolution of the spacecraft orbit. A larger number of Southern Hemisphere samples are taken at larger radial distances. This causes some challenges for our north-south comparison which need to be addressed for an unbiased comparison.

Finally, Figure 2c shows lobe densities in the two hemispheres as a function of radial distance. Densities do not vary significantly with distance tailward of ≃10 $R_E$, but there seems to be a north-south asymmetry in the density. Northern lobe densities are higher than the corresponding southern lobe densities. For the full data set, this asymmetry is present over the full range of radial distances tested; Northern Hemisphere values are consistently higher. For the equinox data set, the difference is still present, albeit smaller. Tailward of approximately 14 $R_E$, northern lobe values seem to indicate a trend toward increasing density with radial distance. A similar feature, though less pronounce, is also seen for the Southern Hemisphere tailward of 17 $R_E$. This is an artifact of the orbit; these samples are taken closer to the plasma sheet, where we expect higher densities.

## 3. Analysis

### 3.1. Factors Influencing Lobe Density

Before we investigate asymmetries in detail, let us just briefly recapitulate some of the most important processes and factors that can potentially influence plasma density in the lobes. We will refer to this list in the discussion section.

*Solar wind dynamic pressure.* An increased solar wind dynamic pressure will compress the whole magnetosphere. Consequently, the existing plasma is squeezed into a smaller volume. This will be reflected as a higher density. On the other hand, a compression also leads to some heating and thus higher plasma temperatures. Noting that the methodology used in the present paper is limited to cold plasma, some of the increase due to compression may not be detected.

*Solar illumination.* In the polar cap region, ionization is primarily driven by solar irradiance in the extreme ultraviolet (EUV) wavelength range. Enhanced solar activity typically leads to higher EUV radiation which in turn will lead to higher ion outflow and thus enhanced lobe density. In addition to the long-time period solar cycle variability, there is of course a much stronger diurnal and seasonal variation in illumination.

*Geomagnetic disturbance level.* Higher geomagnetic activity is often associated with dayside reconnection and thus faster plasma convection and enhanced transport of plasma from the ionosphere. Transient processes such as substorms and bursty bulk flow activity [*Angelopoulos et al.*, 1992] unload the lobe [e.g., *Caan et al.*, 1975]. Both of these processes change the lobe density, but once again heating may shift part of the plasma population out of range of the methodology used here. Traveltimes for cold plasma from the high-latitude ionosphere (source region) to the lobes are several hours. Auroral activity enhances outflow, but travel paths of the more energetic outflow from auroral latitudes do not necessarily thread the lobe region we observe. This means that one would not necessarily expect any strong or direct correlation between lobe density and, e.g., auroral activity.

*Radial distance.* The cross section of magnetic flux tubes increases with increasing radial distance from Earth. Consequently, plasma density falls off with increasing radial distance. Furthermore, the density is higher close to the plasma sheet and close to the magnetopause.

One should have in mind that these factors are not totally uncorrelated. For example, during solar maximum, both solar irradiance and geomagnetic activity increases.

### 3.2. North-South Asymmetries

To further characterize and quantify any north-south asymmetries, we select reference volumes in the Northern and Southern Hemispheres and extract data sampled within these volumes only. As illustrated in Figure 3, these reference volumes only encompass a small fraction of the space sampled by Cluster over the 16 years, and we only use measurements satisfying our lobe critera in Table 1. We then produce distributions similar to Figure 2a and calculate statistical moments (mean, mode, median, and standard deviation) from these subsetted distributions.

We tried various approaches to eliminate any bias due to spacecraft orbit, data coverage, solar wind conditions, and disturbance levels. Note that with our procedure we effectively select data records which have been sampled at different times and thus under different solar wind conditions and geomagnetic activity levels. Ideally, these external factors should be identical for the Northern Hemisphere and Southern Hemisphere data sets to get an unbiased comparison. In reality, however, it is not possible to pick out data subsets where all these external factors are identical in the Northern and Southern Hemispheres. When selecting reference volumes for comparison, we therefore tried various approaches to minimize any bias.

*Comparing densities within fixed volumes in the northern and southern lobes.* A first, rough estimate of any asymmetries can be obtained by comparing the density inside a fixed volume in the northern lobe with the density in an identical volume in the southern lobe. To do this, we define a rectangular box of a certain size, centered in each lobe, and calculate averages within this volume as illustrated in Figure 3a. Seasonal variations can be minimized by selecting data from a narrow (in dawn-dusk direction) volume around magnetic midnight—a region traversed by Cluster around equinox. We tried several volume sizes for our comparison to ensure optimal data coverage (large volumes desirable) and at the same time avoid orbital and seasonal biases. In the initial years of the Cluster mission, the line of apsides of the orbit was roughly in the ecliptic plane, and the data coverage is roughly equal in the Northern and Southern hemispheres. The radial distribution of the measurements is also roughly the same, and there is thus no significant orbital bias in the data. For later years, the line of apsides drops down, and the distribution of measurement location becomes skewed in our fixed volume: There are relatively more samples closer to Earth and closer to the plasma sheet in the Northern Hemisphere, and vice versa for the Southern Hemisphere (see indicated orbits in Figure 1 and the plot of density versus radial distance in Figure 2b).

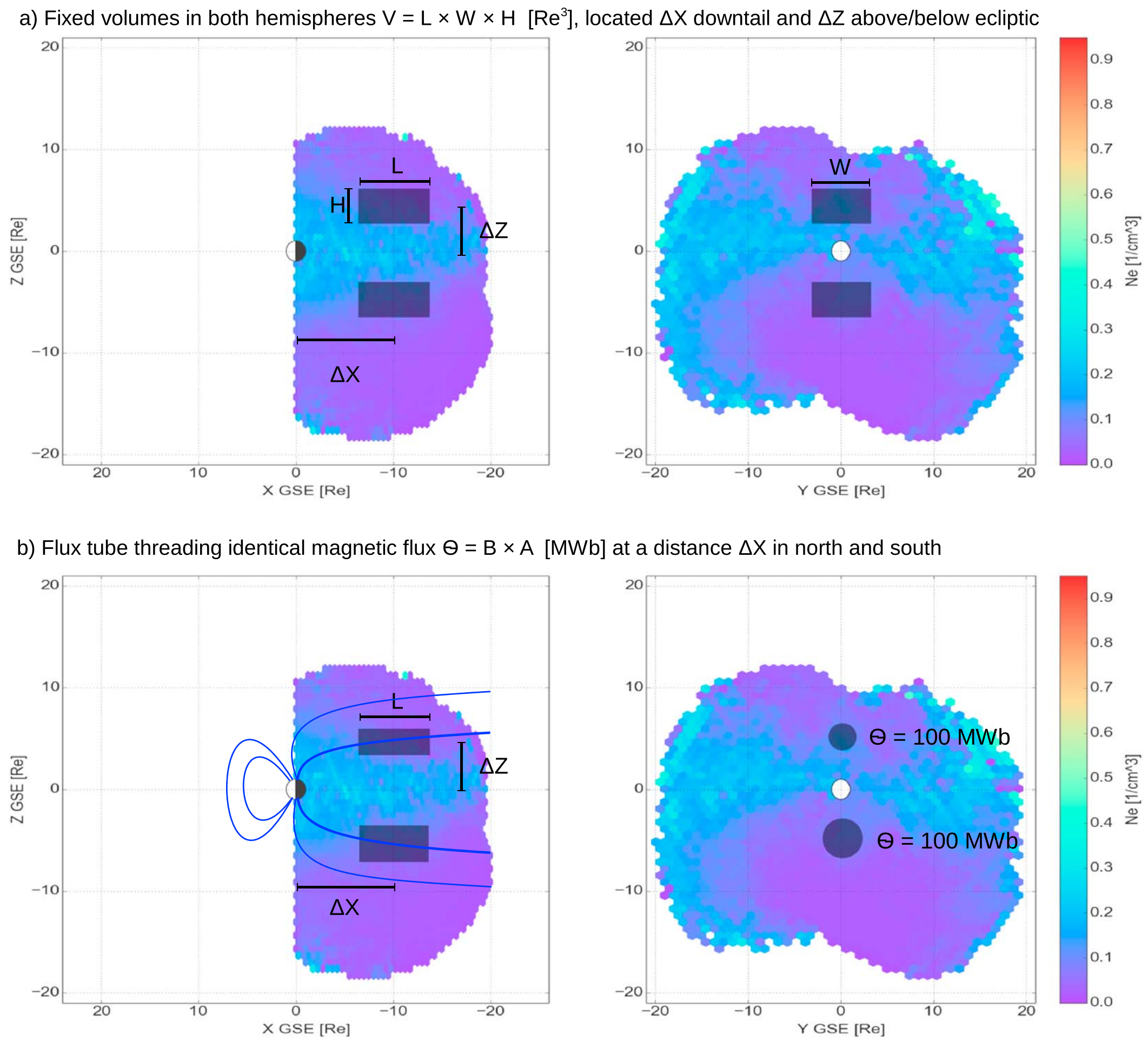

**Figure 3.** Definition of reference volumes in northern and southern lobes used for comparison. The color-coded background image is similar to Figure 1, but now only showing measurements satisfying our lobe criteria in Table 1. (a) Comparison of two equally large rectangular volumes in northern and southern lobe. (b) Reference volumes based on flux tube content. Each flux tube contains 100 MWb magnetic flux and map to the polar cap ionosphere. The average B field in the Northern and Southern Hemispheres can differ, so the two flux tubes may have different cross section (though this is exaggerated here for illustrative purposes). To create sample distributions with the similar properties, we also move and expand/contract the southern flux tube.

*Comparing densities within flux tube segments.* A more refined approach is to define the reference volume from the magnetic flux rather than a fixed volume. The rationale is that ions emanating from the ionosphere are confined within flux tubes as they escape into the magnetospheric lobes, and that the magnetic field may be different in the northern and southern lobes. We can also move the volumes in radial direction to ensure similar radial distribution of the measurements in the two hemispheres. This approach is illustrated in the two panels of Figure 3b. The flux tube is defined as follows.

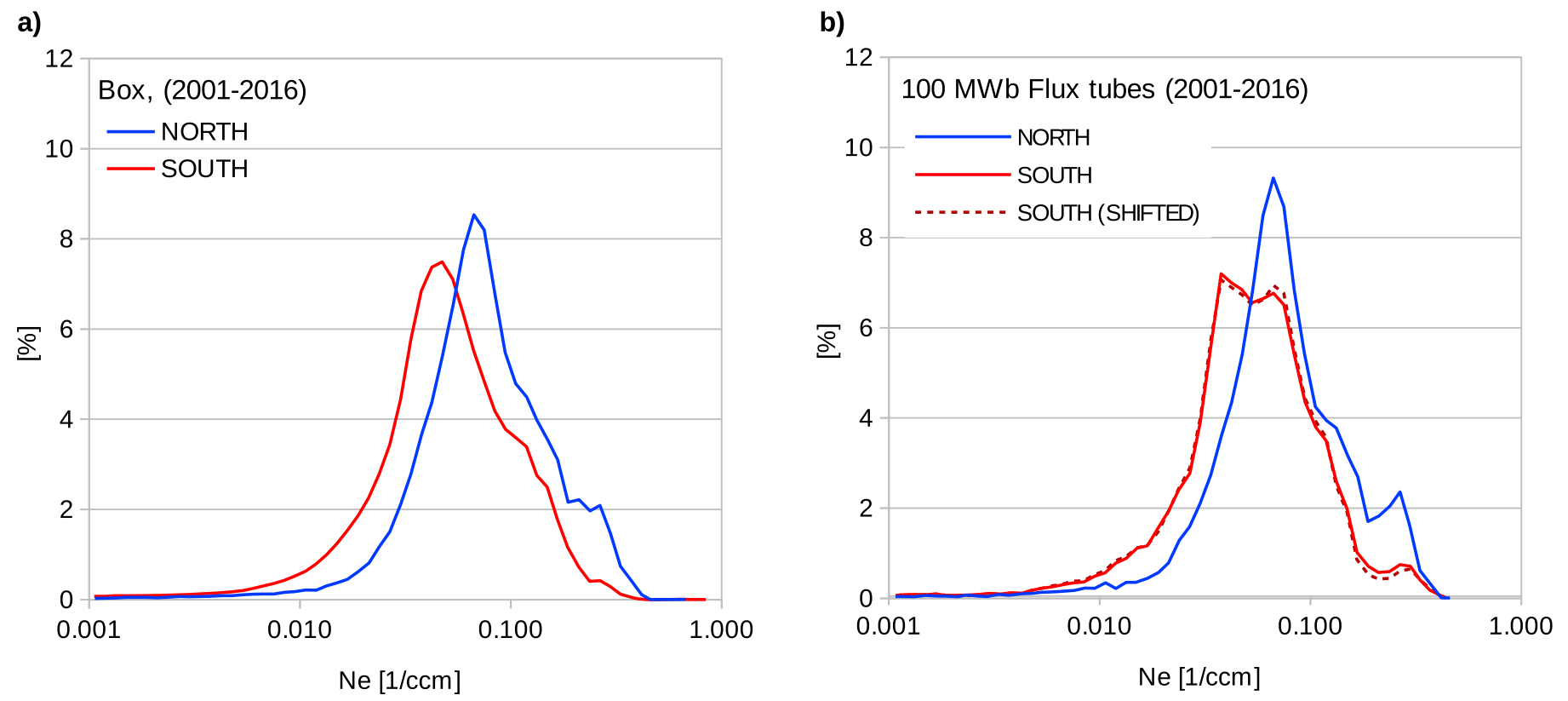

**Figure 4.** Distributions similar to Figure 2a but now only containing measurements inside the reference volumes. (a) Distributions within the rectangular boxes. (b) Distributions within the 100 MWb flux tubes.

1. We define a surface, $S$, in the northern lobe $YZ_{GSE}$ plane at $X = -10R_E$ downtail with a total magnetic flux, $\Theta = \int \vec{B} \cdot \vec{dS} = 100$ MWb. The magnetic field, $B$, measured by Cluster, is used for this calculation.
2. Next, we use a length segment, $L$, to define a flux tube located at distance $\Delta Z$ above the ecliptic in the Northern Hemisphere lobe. The flux tube is centered around magnetic midnight. This region is traversed by Cluster around equinox, so there should not be any significant seasonal bias.
3. We place a similar tube volume at the same $Z$ distance in the southern lobe. Since the actual magnetic field may be different in the southern lobe, the cross section of the flux tube in units of ($m^2$) and consequently the flux tube volume in units of ($m^3$) will, in general, be different from the Northern Hemisphere reference volume.

The size of 100 MWb in step (1) and length segments of the flux tubes are once again compromises to get sufficient statistics (larger volume is better) and at the same time encompass a volume reasonably covered by Cluster. At $\Delta X = 10\ R_E$ downtail, 100 MWb corresponds to a flux tube radius of around 3–4 $R_E$. For comparison, the polar cap area, to which the lobes are magnetically connected, is typically in the range 400–1200 MWb for each hemisphere [e.g., *Sotirelis et al.*, 1998; *Milan et al.*, 2008].

In an attempt to eliminate external factors, we also selectively discarded individual measurements in the Southern Hemisphere volumes to achieve nearly identical distributions of solar wind and geomagnetic activity parameters. This exercise proved to be nearly impossible with our trial-and-error scheme (i.e., essentially tweaking parameters in the database query scripts) and was eventually abandoned.

## 4. Results

A synthesized overview of results are presented in Table 2. Plots of distributions for the full (2001–2016) data set inside the reference volumes are shown in Figure 4. In the table, we show average (median) densities and their standard errors for both hemispheres. We also show averages of some of the key external factors that may influence the lobe density (see section 3.1). Note that the listed auroral electrojet index (*AE*), disturbed storm time index (*Dst*), $F_{10.7}$, and solar wind dynamic pressure ($P_{dyn}$) values are averages over the time intervals when Cluster was actually taking measurements in the lobes—not yearly averages. For convenience, rows are numbered 1 to 12 and columns are labeled A to Q. We will use this chessboard style reference in the discussion below. For each method tested, we calculated characteristic moments for the full data set (2001–2016) and for shorter intervals. In the table we only give median values, since this moment (unlike mode) is unique for each distribution.

The period 2001–2003 is close to solar maximum; 2004–2006 roughly represents intermediate solar activity; 2007–2009 is close to solar minimum. The $F_{10.7}$ index—a proxy for solar activity—given in columns D (average $F_{10.7}$ over times when Cluster sampled the northern lobe reference volume) and L (southern lobe) reflects this change in solar activity. Due to the orbit evolution, we end up with very few data points in the northern lobe during later years, and these samples are also take closer to the plasma sheet. Data from later years are thus less suitable to characterize asymmetries, and we do not present moments for individual intervals after 2009 at all. Column Q shows the ratio between northern and southern lobes average densities. We immediately note that northern lobe cold plasma density values are consistently higher than corresponding southern lobe densities when comparing the results for a specific method and using the same time interval. Below, we discuss some of the individual entries in the table in the context of the processes listed in section 3.1.

**Table 2.** Average (Median) Densities (Columns H and O for Northern and Southern Lobes, Respectively), Their Standard Errors (Columns I and P), and Auxiliary Parameters[a]

| A | B | C | D | E | F | G | H | I | J | K | L | M | N | O | P | Q |
|---|---|---|---|---|---|---|---|---|---|---|---|---|---|---|---|---|
| | | Northern Lobe | | | | | | | Southern Lobe | | | | | | | |
| | Interval (years) | Obs (h) | *AE* (nT) | *Dst* (nT) | $F_{10.7}$ (sfu) | $P_{dyn}$ (nPa) | *N* ($cm^{-3}$) | $\delta N$ ($cm^{-3}$) | Obs (h) | *AE* (nT) | *Dst* (nT) | $F_{10.7}$ (sfu) | $P_{dyn}$ (nPa) | *N* ($cm^{-3}$) | $\delta N$ ($cm^{-3}$) | $\frac{N_{north}}{N_{south}}$ |
| *Exp. 1: Rectangular Box 12 × 12 × 10 [L,W,H $R_E$] at $\Delta X = 12\ R_E$ and $\Delta Z = 6.5\ R_E$ Above/Below Ecliptic* | | | | | | | | | | | | | | | | |
| 1 | 2001–2016 | 2,336 | 124 | −18 | 110 | 1.4 | 0.0723 | 0.0002 | 15,829 | 87 | −7 | 94 | 1.1 | 0.0493 | 0.0000 | 1.46 |
| 2 | 2001–2003 | 1,148 | 175 | −23 | 164 | 1.6 | 0.0787 | 0.0003 | 2,071 | 257 | −17 | 138 | 1.6 | 0.0634 | 0.0002 | 1.24 |
| 3 | 2004–2006 | 1,117 | 94 | −14 | 89 | 1.3 | 0.0647 | 0.0002 | 3,369 | 128 | −16 | 91 | 1.2 | 0.0440 | 0.0001 | 1.47 |
| 4 | 2007–2009 | 57 | 62 | −5 | 68 | 1.2 | 0.0688 | 0.0006 | 3,635 | 45 | −3 | 69 | 1.1 | 0.0405 | 0.0001 | 1.69 |
| *Exp.2: 100 MWb Flux Tube Segments at $\Delta X = 10 \pm 1\ R_E$ and $\Delta Z = 6.5\ R_E$ Above/Below Ecliptic* | | | | | | | | | | | | | | | | |
| 5 | 2001–2016 | 305 | 141 | −16 | 109 | 1.3 | 0.0700 | 0.0005 | 782 | 97 | −8 | 94 | 1.2 | 0.0526 | 0.0002 | 1.33 |
| 6 | 2001–2003 | 139 | 187 | −17 | 137 | 1.4 | 0.0769 | 0.0008 | 166 | 304 | −16 | 148 | 1.7 | 0.0714 | 0.0007 | 1.07 |
| 7 | 2004–2006 | 162 | 91 | −15 | 89 | 1.2 | 0.0638 | 0.0006 | 101 | 100 | −18 | 102 | 1.1 | 0.0591 | 0.0008 | 1.08 |
| 8 | 2007–2009 | 2 | 19 | −4 | 68 | 0.8 | 0.0765 | 0.0026 | 206 | 54 | −5 | 69 | 1.1 | 0.0386 | 0.0002 | 1.98 |
| *Exp. 3: As Above but for Southern Hemisphere Flux Tube Shifted Closer to Earth* | | | | | | | | | | | | | | | | |
| 9 | 2001–2016 | 305 | 141 | −16 | 109 | 1.3 | 0.0700 | 0.0005 | 646 | 85 | −8 | 89 | 1.1 | 0.0521 | 0.0002 | 1.34 |
| 10 | 2001–2003 | 139 | 187 | −17 | 137 | 1.4 | 0.0769 | 0.0008 | 147 | 309 | −15 | 150 | 1.7 | 0.0679 | 0.0007 | 1.13 |
| 11 | 2004–2006 | 162 | 91 | −15 | 89 | 1.2 | 0.0638 | 0.0006 | 79 | 117 | −16 | 94 | 1.2 | 0.0563 | 0.0008 | 1.13 |
| 12 | 2007–2009 | 2 | 19 | −4 | 68 | 0.8 | 0.0765 | 0.0026 | 231 | 54 | −4 | 69 | 1.0 | 0.0402 | 0.0003 | 1.90 |

[a]Blocks of rows indicate different time intervals and methods. sfu, solar flux unit (1 sfu = $10^{-22}$ W m$^{-2}$ Hz$^{-1}$). See text for details.

Rows 1–4 summarize our experiment with fixed rectangular volumes. Here the 2001–2003 time interval (row 2), i.e., in the initial years of the Cluster mission, is least affected by orbital bias. Average radial distance of sampling locations are more similar than later years. One could argue, with some credibility, that there is a bias caused by external conditions. The average solar illumination (reflected by $F_{10.7}$, given in column F) and also the general geomagnetic background activity (reflected in the *Dst* index shown in column E) are higher during periods Cluster sampled the northern lobe. This argument does not hold for later years, though.

In rows 5–8, we show the corresponding averages from our flux tube experiment. Several flux tube length segments (see Figure 3) were tested, but in this table we only show the results obtained with an $L = 2\ R_E$ long flux tube segment. We thus include samples from much smaller volumes than the rectangular boxes above, but the effect of orbital bias is reduced since we now probe a shorter radial distance range and smaller volumes. Once again, northern lobe densities are consistently higher than southern lobe densities. For the full data set (2001–2016, shown in row 5), one could again argue that differences in external conditions are the cause of the asymmetry. For the individual time ranges (rows 6–8), this argument does not hold, though. Here both the average solar wind dynamic pressure and the average $F_{10.7}$ are higher for the southern lobe samples. One would actually expect higher densities in the southern lobe if this argument was to be used, but this is not the case. The results still show consistently higher northern lobe density values.

Finally, rows 9-12 show results obtained when we tried to achieve identical distributions of the sampling locations, i.e., we try to completely remove any remaining orbital bias. The Northern Hemisphere reference volume was kept as it is, so these numbers are identical to those in rows 5–8, but the Southern Hemisphere reference volume was moved (up to a few $R_E$) inward and the cross section adjusted so that the 100 MWb total flux was maintained. We did not achieve perfectly identical distributions of radial locations in the northern and southern lobes, but the mean and median values of the sampling locations were within a few 100 km apart for Northern and Southern Hemispheres. Using this procedure, which is probably the best in terms of removing any bias, still shows a north-south asymmetry.

We also tried to achieve similar average $Z$ distance for the northern and southern distributions by varying the $\Delta Z$ or limiting the height of the rectangular boxes. There is no way of knowing the instantaneous position or thickness of the plasma sheet for each measurement in our data set. However, around equinox, a reasonable assumption is that the plasma sheet is on average centered about the ecliptic plane ($Z \simeq 0$), so that our $\Delta Z$ (see Figure 3) can be used as proxy for the distance to the central plasma sheet. The asymmetry still exists if enforces similar $\Delta Z$ distributions for the northern and southern lobe data sets, but this is only possible for the full data set and the 2001–2003 subsets.

It is of course possible to set up the experiment so that southern lobe densities become equal or even higher than northern lobe densities, for example, by moving the southern lobe reference volume much closer to Earth or much closer to the plasma sheet. With our data set, this could only be achieved with unrealistic placements of the Southern Hemisphere reference volume, though.

The north-south asymmetries in the above results are fairly robust and do not depend strongly on our definition of lobes (see Table 1) or size of reference volumes. We tried to increase the magnetic field requirements, requiring minimum field strengths of 10 nT or 20 nT. The overall results remained similar. Also, neither changes in the thresholds for plasma beta, or even completely discarding the beta threshold completely, changed the overall results significantly. The asymmetry also remains if we remove the high density part of the distribution by requiring $N_{\mathrm{lobe}} \leq 0.1$ cm$^{-3}$. This criteria, though perhaps not justified, should remove any remaining central plasma sheet and plasma sheet boundary layer contributions.

## 5. Discussion and Summary

In terms of average densities and dependence on solar illumination, the new, extended Cluster cold plasma data set does not differ much from values reported earlier [e.g., *Escoubet et al.*, 1997b; *Laakso et al.*, 2002; *Svenes et al.*, 2008; *Haaland et al.*, 2012]. Typical plasma densities are below 0.1 cm$^{-3}$ in the lobes.

Given the strong seasonal variation in EUV illumination and ion outflow, a north-south asymmetry in lobe density is perhaps not surprising. More interesting is that the experimental results from Cluster also indicate a north-south asymmetry in cold plasma density around equinox. Observed northern lobe density values, regardless of which statistical moment we use, are higher than southern lobe density values. Although the methodology used to obtain the results— inferring cold plasma density from spacecraft potential—has a finite accuracy, it should not influence the asymmetry. There is significant spread in the statistical data, but due to the large number of data points, the standard error is typically very low. Using 16 years of data and checking several moments of the distributions should ensure that strong effects caused by single events or outliers in the data are minimized. Some bias due to the spacecraft orbit and sampling of data may still exist, but in the analysis of the measurements, we have tried various approaches to eliminate bias in the data. The persistent asymmetry inferred thus seems to be real.

Much of the explanations for the persistent north-south asymmetry in lobe density observed can probably be found in properties of the high-latitude ionosphere—the dominant source of cold plasma to the Earth's magnetosphere. In fact, a north-south asymmetry should not come as a surprise. On the contrary, given the different properties of the ionosphere in the two hemispheres, of which some are illustrated in Figure 5, a perfectly symmetric magnetosphere in terms of lobe density would not be expected.

A detailed, quantitative assessment of the individual contributions from each of the parameters illustrated in Figure 5 is beyond the scope of the present paper and will be the focus of a later companion paper. Here we just briefly point out some of the main differences between the Northern and Southern Hemisphere and try to relate them to the observed lobe asymmetry.

First, the magnetic field configuration and strength are different in the Northern and Southern Hemispheres as illustrated in Figure 5a. Poleward of 80° corrected geomagnetic latitude, the magnetic field at low altitudes in the south is stronger by 7% on average [e.g., *Laundal and Richmond*, 2016]. The magnetic field influences conductivities and currents in the ionosphere, and also scale heights related to ionization and outflow [see e.g., *Cnossen et al.*, 2011; *Cnossen*, 2017]. Since, at least on average, the total amount of magnetic flux in the Northern and Southern Hemispheres must be equal, the polar cap area (defined by open magnetic field lines) and thus source area for ion outflow will thus be different between the Northern and Southern Hemispheres (Figure 5b). Likewise, the larger offset between the magnetic and geographic poles in the Southern Hemisphere leads to larger diurnal variations in solar illumination in the Southern Hemisphere [e.g., see *Barakat et al.*, 2015; *Maes et al.*, 2016] (Figure 5c).

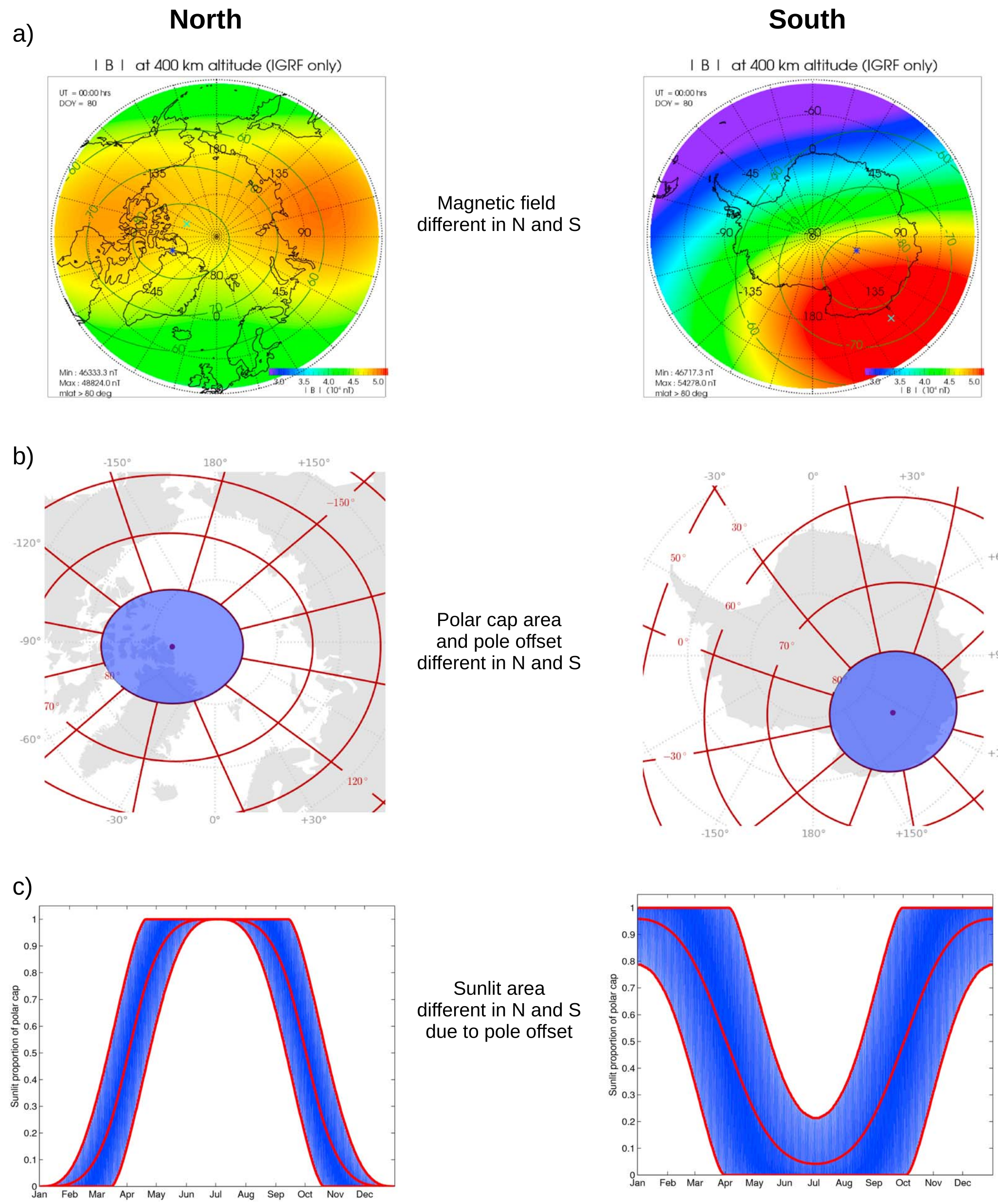

**Figure 5.** Possible explanations for north-south differences. (a) Differences in the magnetic field [after *Förster and Cnossen*, 2013]. (b) Differences in the polar cap shape, area, and pole offset. Polar cap area above 80° corrected geomagnetic latitude is highlighted [after *Laundal and Richmond*, 2016]. (c) Differences in annual illumination of the polar cap area. Horizontal axes show season; vertical axes show the fraction of polar cap area illuminated assuming a terminator at 90° solar zenith angle. [after *Maes et al.*, 2016].

Although the measurements presented here are taken around equinox, the exobase height and scale height are not the same in the Northern and Southern Hemispheres, despite the same illumination. There is a phase lag of around 2 months, so spring is more "winter like," and autumn is more "summer like" [*Cnossen and Förster*, 2016]. The effect of this thermospheric time lag on ion outflow and magnetospheric dynamics remains elusive though.

The importance of the ionosphere and ion outflow is also noted in the long-time (solar cycle) variation of the lobe density. Periods with high solar activity and thus high EUV irradiance are characterized by higher lobe densities than periods with low solar activity as already noted in *Svenes et al.* [2008] and *Haaland et al.* [2012]. This long-term variation does not seem to have any fundamental influence on the north-south asymmetry, though.

From the observations, we also note that the magnetic flux tubes have slightly larger cross sections in the southern lobe. This means that even with identical ionospheric supply, the outflow would be distributed into a larger flux tube volumes in the southern lobe, which would mean that the density there would be lower.

The Cluster data set used in the present study is quite unique in terms of coverage and accuracy, so a comparison to other studies is difficult. Very preliminary examinations of other Cluster data sets also indicate some north-south asymmetries, though. The authors have access to the cold ion outflow data set used by *Engwall et al.* [2009] and a more recent and more comprehensive data set presented in *André et al.* [2015]. These data sets focus on the field-aligned transport of cold ions, using spacecraft wake measurements [e.g., *Engwall et al.*, 2006] to infer outflow velocities combined with spacecraft charging measurements to obtain densities needed to estimate fluxes. A more quantitative study of asymmetries in these data sets will also be the subject of a future paper. Model and simulation results focusing on hemispheric asymmetries in outflow are scarce, but a recent study by *Barakat et al.* [2015] demonstrated that north-south asymmetries in outflow can be reproduced if realistic boundary conditions are used to parametrize models.

Finally, one may ask what the implications of an asymmetry in lobe density might be. Provided that the plasma temperature, $T$, is the same in northern and southern lobes, the thermal pressure, $p = nkT$, where $k$ is the Boltzmann constant, will be different. However, as indicated above, Southern Hemisphere flux tubes are larger, implying a weaker magnetic field. Plasma temperatures may also be different, but these can only be estimated from particle distributions, which are difficult to determine with accuracy in the lobes. The lower density in one lobe therefore does not necessarily imply a system in imbalance.

**Acknowledgments**
Data from the Cluster satellites and auxiliary data were obtained from the Cluster Science Archive (http://www.cosmos.esa.int/web/csa). Computer code used for the calculations in this paper has been made available as part of the QSAS science analysis system. QSAS is provided by the United Kingdom Cluster Science Centre (Imperial College London and Queen Mary, University of London) supported by the United Kingdom Science and Technology Facilities Council (STFC). Solar wind data were obtained from the Coordinated Data Analysis Web (CDAWeb — see http://cdaweb.gsfc.nasa.gov/about.html). We also thank the International Space Science Institute, Bern, Switzerland for providing computer resources and infrastructure for data exchange for ISSI team 308 — "Magnetosphere-ionosphere-thermosphere Coupling: Differences and Similarities between the Two Hemispheres". This work was supported by the Research Council of Norway/CoE under contract 223252/F50 and Deutsches Zentrum für Luft- und Raumfahrt, grant 50OC1401. L.M. acknowledges support from BELSPO IAP PLANET TOPERS (P7/15).